\documentclass[a4paper,12pt,eqno]{article}
\usepackage{theorem}
\usepackage{latexsym,amssymb,amsfonts,amsmath}
\usepackage{graphicx}
\newtheorem{thm}{Theorem}[section]
\newtheorem{lem}[thm]{Lemma}

\newtheorem{pro}[thm]{Proposition}
\newtheorem{rem}[thm]{Remark}

{\theorembodyfont{\upshape}
\newtheorem{defi}[thm]{Definition}

\theoremheaderfont{\scshape}
\newcommand{\RM}{\mathbb{R}}
\newcommand{\ZM}{\mathbb{Z}}
\newcommand{\ZMP}{\mathbb{Z}_{+}}

\newcommand{\NM}{\mathbb{N}}
\newcommand{\CM}{\mathbb{C}}

\newcommand{\EM}{\mathbb{E}}

\def\mcB{\mathcal{B}}

\def\mcE{\mathcal{E}}

\def\mcH{\mathcal{H}}
\def\mcK{\mathcal{K}}
\def\mcL{\mathcal{L}}
\def\mcM{\mathcal{M}}

\newenvironment{pf}{\bigskip\par\noindent{\it Proof:}}%
                    {$\square$\bigskip}

\newcommand{\ket}[1]{|#1 \rangle}

\title{{\Large {\bf Limit theorems for open quantum random walks}}
\author{
{\small Norio Konno}\\
{\scriptsize Department of Applied Mathematics, 
Faculty of Engineering, 
Yokohama National University}\\
{\scriptsize 79-5 Tokiwadai, Hodogaya, Yokohama, 240-8501, Japan}\\
{\scriptsize e-mail: konno@ynu.ac.jp, Tel.: +81-45-339-4205, Fax: +81-45-339-4205}\\
{\small Hyun Jae Yoo}\\
{\scriptsize Department of Applied Mathematics,
Hankyong National University}\\
{\scriptsize 327 Jungangro, Anseong-si, Gyeonggi-do, 456-749, Korea}\\
{\scriptsize e-mail: yoohj@hknu.ac.kr, Tel.: +82-31-670-5344, Fax: +82-31-670-5349}\\}
}
\vskip 1cm

\date{\empty }
\pagestyle{plain}

\begin{document}
\maketitle

\par\noindent
\begin{small}
\par\noindent
{\bf Abstract}. We consider the limit distributions of open quantum random walks on one-dimensional lattice space. We introduce a dual process to the original quantum walk process, which is quite similar to the relation of Schr\"odinger-Heisenberg representation in quantum mechanics. By this, we can compute the distribution of the open quantum random walks concretely for many examples and thereby we can also obtain the limit distributions of them. In particular, it is possible to get rid of the initial state when we consider the evolution of the walk, it appears only in the last step of the computation. 
  
\footnote[0]{
{\it Abbr. title:} Dynamics for open quantum random walks
}
\end{small}

\setcounter{equation}{0}
\section{Introduction}
Recently Attal et al. \cite{AttalEtAl2011a, AttalEtAl2011b} introduced and investigated the open quantum random walk (OQRW) on graphs, which shows various different dynamical behaviors comparing to usual quantum walks and it includes the classical random walk as a special case. Then they considered the limit theorems for OQRW's: they have shown the central limit theorem \cite{AG-PS2012}. The purpose of this paper is to further investigate the distribution and the limit theorem of OQRW's.

A quantum analog of the classical random walk, called quantum walk (QW), has been intensively studied for the last decade (see \cite{Kempe2003, Kendon2007, Konno2008, VAndraca2008}). The most remarkable difference between classical random walk and quantum walk appears in the central limit theorem. In the classical random walk, the limit distribution is Gaussian with scaling speed $\sqrt{n}$. On the other hand, in quantum walk the speed is linear in time, i.e., $n$, and moreover, the limit distribution is far from Gaussian \cite{ABNVW2001, GJS2004, KY2012, Konno2002, Konno2005}. The main reason that makes quantum walk different from classical walk is the interference.

The OQRW is different from the usual discrete-time QW. It was introduced in order to model the quantum efficiency in biological systems and quantum computing and it is based on the non-unitary dynamics induced by the local environments \cite{AttalEtAl2011a, AttalEtAl2011b}. These random walks deal with density matrices instead of pure states. In \cite{AttalEtAl2011b}, Attal et al. developed the quantum trajectory approach for OQRW and by using this concept, they have shown the central limit theorem for OQRW's on  the $d$-dimensional integer space $\ZM^d$ \cite{AG-PS2012}. 

In this paper we focus on OQRWs on $\ZM$. We will introduce a concept of dual process to the original OQRW.  By this we may think that the OQRW is a process where the environment evolves rather than the walker itself evolves as time goes on. This is particularly useful when we compute the distribution of the walk because the initial states are not relevant during the evolution. See section 2 for the details. Moreover, for many examples, we can compute the distribution of the walk very concretely and thereby we can also get the limit distributions of them.  This paper is organized as follows. In section 2, we introduce the concept of OQRW following \cite{AttalEtAl2011b} and new concept of dual process. Next we give our main result. Section 3 is devoted to the proof of our main result and some preparation that is useful for examples. In section 4, we consider several concrete examples.

\section{Open Quantum Random Walks}
In this section, we give a brief definition of OQRW and define a dual process of it. Then we state the main result. In order to compare, we first shortly review the usual quantum walks, so called unitary quantum walks.

\subsection{Unitary Quantum Walks} The discrete-time QW is a quantum version of the classical random walk with an additional degree of freedom called chirality. The chirality takes values left and right, and it determines the direction of the motion of the walker. At each time step, if the walker has the left chirality, it moves one step to the left, and if it has the right chirality, it moves one step to the right. In this paper, we put
\begin{eqnarray*}
\ket{L} = 
\left[
\begin{array}{cc}
1 \\
0  
\end{array}
\right],
\qquad
\ket{R} = 
\left[
\begin{array}{cc}
0 \\
1  
\end{array}
\right],
\end{eqnarray*}
where $L$ and $R$ refer to the left and right chirality state, respectively.  

For the general setting, the time evolution of the walk is determined by a $2 \times 2$ unitary matrix, $U$, where
\begin{align*}
U =
\left[
\begin{array}{cc}
a & b \\
c & d
\end{array}
\right],
\end{align*}
with $a, b, c, d \in \mathbb C$ and $\CM$ is the set of complex numbers. The matrix $U$ rotates the chirality before the displacement, which defines the dynamics of the walk. To describe the evolution of our model, we divide $U$ into two matrices:
\begin{eqnarray*}
P =
\left[
\begin{array}{cc}
a & b \\
0 & 0 
\end{array}
\right], 
\quad
Q =
\left[
\begin{array}{cc}
0 & 0 \\
c & d 
\end{array}
\right]
\end{eqnarray*}
with $U = P + Q$. The important point is that $P$ (resp. $Q$) makes the walker moves to the left (resp. right) at position $x$ at each time step. For example, the {\it Hadamard walk} is determined by the Hadamard gate $U = H$:
\begin{eqnarray*}
H=\frac{1}{\sqrt2}
\left[
\begin{array}{cc}
1 & 1 \\
1 &-1 
\end{array}
\right].
\end{eqnarray*}
The walk is intensively investigated in the study of the QW.

Let $\Xi_{n} (l,m)$ denote the sum of all paths starting from the origin in the trajectory consisting of $l$ steps left and $m$ steps right at time $n$ with $n=l+m$. For example, 
\begin{align*}
\Xi_2 (1,1) &= Q P + P Q, \\
\Xi_4 (2,2) &= Q^2 P^2 + P^2 Q^2 + Q P Q P + P Q P Q + P Q^2 P + Q P^2 Q. 
\end{align*}
Let $\ZMP = \{0,1,2, \ldots \}.$ The probability that our quantum walker is in position $x \in \ZM$ at time $n \in \ZMP$ starting from the origin with $\varphi = {}^t [\alpha, \beta]$ with $\alpha, \> \beta \in \CM$ and $|\alpha|^2 + |\beta|^2 = 1$ is defined by 
\begin{align}\label{eq:probability_density}
P (X_{n} =x) = \| \Xi_{n}(l, m) \> \varphi \|^2,
\end{align}
with $n=l+m$ and $x=-l+m$ where ${}^t [\alpha, \beta]$ means the transpose of $[\alpha, \beta]$.

The reason that we call this quantum walk the unitary quantum walk is that the evolution of the walk is the unitary transform of a state in a Hilbert space. To say more concretely, let us denote by $\mcH_C:=\CM^2$
the space of intrinsic structure, namely the chirality and let 
$\mcK_S:=l^2(\ZM)$ the space of positions. The Hilbert space on which the quantum walk evolves is given by 
\[
\mcH:=\mcH_C\otimes\mcH_S\thickapprox l^2(\ZM, \CM^2)
\]
and any state, i.e., a unit vector of $\mcH $ is given by 
\[
\psi=(\psi(x))_{x\in \ZM}, \quad \psi(x)\in \CM^2.
\] Let us denote by $T$ the left translation in $l^2(\ZM)$:
\[
(Ta)(x)=a(x+1), \quad \text{for }a=(a(x))_{x\in \ZM}.
\]
$T$ is a unitary map whose adjoint is the right translation:
\[
(T^*a)(x)=a(x-1), \quad \text{for }a=(a(x))_{x\in \ZM}.
\]
Let $\psi_0\in \mcH $ be any initial state. Then the dynamics of $1$-dimensional quantum walk driven by the unitary operator $U=P+Q$ in the above is represented as
\[
\psi_n=(PT+QT^*)^n\psi_0,
\]
where $\psi_n=(\psi_n(x))_{x\in \ZM}\in \mcH$ is the state at time $n$. It is easy to see that the operator $PT+QT^*$ is a unitary operator (here tacitly we understand the operators $P$ and $Q$ are the natural extensions of the original ones to $\oplus_{x\in \ZM}\mcB(\mcH_C)$ and hence the quantum walk is a unitary evolution on the Hilbert space $\mcH $. The probability density of \eqref{eq:probability_density} is then nothing but 
\[
P(X_n=x)=\|\psi_n(x)\|^2,
\]
with initial state $\psi_0=\varphi\otimes |0\rangle$. Here $\{|x\rangle:\,x\in \ZM\}$ denotes the canonical orthonormal basis of $l^2(\ZM)$.  

The limit distribution of quantum walk is very different from that of classical random walk. It is ballistic instead of diffusive in the sense that $X_n/n$ converges weakly to a limit whose density function, rigorously shown by Konno \cite{Konno2002, Konno2005}, is given by (in the case $abcd\neq 0$)
\[
x\mapsto \frac{\sqrt{1-|a|^2}(1-\beta x)}{\pi (1-x^2)\sqrt{|a|^2-x^2}}1_{(-|a|,|a|)}(x),
\]
where $\beta$ is a constant depending on the initial state $\varphi$ and the unitary matrix $U$. Here $1_A (x) = 1 \> (x \in A), \> = 0 \> (x \not\in A).$

\subsection{Open Quantum Random Walks}
In this subsection we introduce OQRW's following \cite{AttalEtAl2011b}. The OQRW's can be defined on any dimensional integer spaces as well as on any graphs, but here we confine ourselves to 1-dimensional space $\ZM$ for simplicity.

As before $\mcH_C=\CM^2$ is the Hilbert space for the intrinsic structure and $\mcH_S=l^2(\ZM)$ for positions. Let $B$ and $C$ be two linear operators on $\mcH_C$, i.e., $2\times 2$ matrices, such that 
\[
B^*B+C^*C=I.
\]
Define a completely positive map on the density matrices of $\mcH_C$ by 
\begin{equation*}\label{cp_map}
\mcL(\rho):=B\rho B^*+C\rho C^*.
\end{equation*}
The OQRW lifts this map to $\mcH=\mcH_C\otimes \mcH_S$ in the following way. We consider density matrices on $\mcH$ of the type: 
\[
\rho=\sum_{x\in \ZM}\rho_x\otimes |x\rangle \langle x|,
\]
where each $\rho_x$ is a positive matrix and satisfy
\[
\sum_{x\in \ZM}\mathrm{Tr}(\rho_x)=1.
\]
Let $\widetilde{\mathcal L}$ be an operator that maps on such density matrices as follows:
\begin{equation}\label{eq:oqrw_map}
\widetilde{\mathcal L}(\rho)=\sum_{x\in\ZM}(B\rho_{x+1}B^*+C\rho_{x-1}C^*)\otimes |x\rangle \langle x|.
\end{equation}
The OQRW is an evolution obtained by iteration of the map $\widetilde{\mathcal L}$. Let $\rho^{(0)}:=\rho_0\otimes |0\rangle\langle0|$ be the initial state, i.e., a density matrix on $\mcH$. Then, the evolution of OQRW on $\ZM$ generated by $B$ and $C$ is given by 
\begin{eqnarray}
\rho^{(n)}&=&\sum_x\rho^{(n)}_x\otimes |x\rangle\langle x|, \label{eq:evolution}\\
\rho^{(n+1)}_x&=&B\rho^{(n)}_{x+1}B^*+C\rho^{(n)}_{x-1}C^*, \quad x\in \ZM, \,\,n=0,1,2,\cdots.\label{eq:evolution1}
\end{eqnarray}
The probability distribution to find out the walker at site $x$ at time $n$ is given by 
\begin{equation}\label{eq:probability_density}
p_x^{(n)}=\mathrm{Tr}(\rho_x^{(n)}), \quad x\in\ZM,\,\,n\ge 0.
\end{equation}

In \cite{AttalEtAl2011b}, Attal et al. introduced the concept of quantum trajectory to OQRW's, from which they could show the central limit theorem for OQRW's \cite{AG-PS2012}. Applied to OQRW's on 1-dimensional space $\ZM$, it says that one can introduce a Markov process on $\mcE(\mcH_C)\times \ZM$, where $\mcE(\mcH_C)$ is the space of density matrices on $\mcH_C$, such that the distribution of the space component of the process coincides with that of OQRW. To say more in detail, it is a Markov chain $(\rho_n, X_n)_{n\in \NM}$ with values in $\mcE(\mcH_C)\times \ZM$ with the following transition rule. From any position $(\rho, X)$ it jumps to one of two states: to $(\frac1{p_B}B\rho B^*, X-1)$ with probability $p_B:=\mathrm{Tr}(B\rho B^*)$ or to $(\frac1{p_C}C\rho C^*, X+1)$ with probability $p_C:=\mathrm{Tr}(C\rho C^*)$. Then the central limit theorem shown in \cite{AG-PS2012} reads as follows (stated in 1-dimensional case).
\begin{thm}[\cite{AG-PS2012}, Theorem 5.2]\label{thm:clt_AG-PS}
Consider the stationary open quantum random walk on $\ZM$ associated to the operators $\{B,C\}$. We assume that the completely positive map 
\begin{equation}\label{eq:cpm}
\mcL(\rho)=B\rho B^*+C\rho C^*
\end{equation}
admits a unique invariant state $\rho_\infty$. Let $(\rho_n,X_n)_{n\ge 0}$ be the quantum trajectory process to this open quantum random walk, then 
\[
\frac{X_n-nm}{\sqrt{n}}
\]
converges in law to the Gaussian distribution $N(0,\sigma^2)$ in $\RM$, with mean 
\[
m=\mathrm{Tr}(C\rho_\infty C^*)-\mathrm{Tr}(B\rho_\infty B^*)
\]
and variance
\[
\sigma^2=\mathrm{Tr}(B\rho_\infty B^*+C\rho_\infty C^*)-m^2 +2\mathrm{Tr}[(C\rho_\infty C^*-B\rho_\infty B^*)L]-2m\mathrm{Tr}(\rho_\infty L),
\]
where $L$ is the solution of the equation
\[
L-\mcL^*(L)=C^*C-B^*B-mI.
\]
\end{thm}
\subsection{Dual Processes and Main Result}
In this subsection we introduce the concept of dual process to the OQRW and  state our main result. Recall that $\mcH_C=\CM^2$ and so $\mcB(\mcH_C)=\mcM_2$, the algebra of all $2\times 2$ matrices. From now on we regard $\mathcal M_2$ as a Hilbert space equipped with the Hilbert-Schmidt inner product:
\begin{equation}\label{eq:H-S_inner_product}
\langle A,B\rangle:=\mathrm{Tr}(A^*B), \quad A,B\in \mathcal M_2.
\end{equation}
Let $\mathcal M:=\oplus_{x\in \ZM}\mathcal M_2$ be the direct sum Hilbert space. Recall the left and right translation operators $T$ and $T^*$ on $l^2(\ZM)$. The operators $T$ and $T^*$ naturally extend to $\mathcal M$. We will use the same symbols whenever there is no danger of confusion. Let $K:=(-\pi,\pi]$ and we understand it as a unit circle on the plane. The Fourier and inverse Fourier transforms between $l^2(\ZM)$ and $L^2(K, \frac1{2\pi}dx)$ are defined as usually: 
\begin{eqnarray*}
\hat a(k)&:=&\sum_{x\in \ZM}e^{-ikx}a(x), \quad a=(a(x))_{x\in \ZM}\in l^2(\ZM),\\
\check f(x)&:=&\frac1{2\pi}\int_{K} e^{ikx}f(k)dk,\quad f\in L^2(K,\frac1{2\pi}dk).
\end{eqnarray*}
For each $k\in K$, let ${\mathcal M}_k$ be the copy of the Hilbert space $\mathcal M_2$ and let 
\[
\hat {\mathcal M}:=\int_{K}^{\oplus}\mathcal M_k\frac1{2\pi}dk
\]
be the direct integral Hilbert space. The Fourier transform also naturally extends to the transform between $\mathcal M$ and $\hat {\mathcal M}$: for $A=(A(x))_{x\in\ZM}\in \mathcal M$, 
\begin{eqnarray*}
\hat A&:=&(\hat A(k))_{k\in K}\in \hat {\mathcal M},\\
\hat A(k)&:=&\sum_{x\in \ZM}e^{-ikx}A(x),
\end{eqnarray*}
and similarly for the inverse transform.

Now let us introduce the left and right multiplication operators on the Hilbert space $\mathcal M_2$. For any $B\in \mathcal M_2$, the left multiplication $L_B$ and the right multiplication $R_B$ are defined on $\mathcal M_2$ by 
\begin{equation}\label{eq:left_right_multiplications}
L_B(A):=BA, \quad R_B(A):=AB, \quad A\in \mathcal M_2.
\end{equation}
Notice that $(L_B)^*=L_{B^*}$ and $(R_B)^*=R_{B^*}$, and for any $B$ and $C$ in $\mathcal M_2$, $L_B$ and $R_C$ commute: $L_BR_C=R_CL_B$. However, $L_B$ and $L_C$ do not commute in general. The operators $L_B$ and $R_B$ are positive definite if $B\ge 0$. Without mentioning further, we will use the same symbols $L_B$ and $R_B$ for the extensions to $\mathcal M$. Thus, as an example, for $A=(A(x))_{x\in \ZM}\in \mathcal M$, 
\[
L_BR_C(A)=(BA(x)C)_{x\in \ZM}\in \mathcal M.
\]

Let us now come back to the OQRW's on $\ZM$. The state $\rho^{(n)}$ at time $n$ can be understood as an element of $\mathcal M$. Then the dynamics \eqref{eq:evolution1} of OQRW's becomes an evolution on $\mathcal M$ given by
\begin{equation}\label{eq:evolution_on_M}
\rho^{(n+1)}=(L_BR_{B^*}T+L_CR_{C^*}T^*)\rho^{(n)}.
\end{equation}
Therefore the solution to \eqref{eq:evolution_on_M} becomes simply
\begin{equation}\label{eq:solution}
\rho^{(n)}=(L_BR_{B^*}T+L_CR_{C^*}T^*)^n\rho^{(0)}.
\end{equation}
If we look at the evolution in the Fourier transform space, then it becomes 
\begin{eqnarray}\label{eq:evolution_in_Fourier_space}
\widehat{\rho^{(n)}}&=&(\widehat{\rho^{(n)}}(k))_{k\in K},\nonumber\\
\widehat{\rho^{(n)}}(k)&=&\left(e^{ik}L_BR_{B^*}+e^{-ik}L_CR_{C^*}\right)^n\widehat{\rho^{(0)}}(k).
\end{eqnarray}
Notice that $\widehat{\rho^{(0)}}(k)$ is the constant (operator valued) function $\rho_0$ because $\rho^{(0)}=\rho_0\otimes |0\rangle\langle0|$ (of course we can take quite general initial state $\rho^{(0)}$ not localized at the origin). 

In order to get the probability distribution, let us define a "dual process":  
\begin{defi}
The dual process to the OQRW generated by $B$ and $C$ is the process $Y_n:=(Y_n(k))_{k\in K}\in \hat{\mathcal M}$ defined by 
\begin{equation}\label{eq:dual_process}
Y_n(k):=\left(e^{ik}L_{B^*}R_B+e^{-ik}L_{C^*}R_C\right)^n(I).
\end{equation}
\end{defi}
The reasoning for the nomenclature becomes clear from the following relation, which is our main result.
\begin{thm}\label{thm:probability_by_dual_process}
The probability distribution of the OQRW is given by 
\begin{equation}\label{eq:probability_by_dual_process}
p_x^{(n)}=\frac1{2\pi}\int_Ke^{ikx}\mathrm{Tr}\left(\rho_0 Y_n(k)\right)dk.
\end{equation}
\end{thm}
\begin{rem}
Notice that in the formula for the distribution of the walker in the above theorem, the initial state $\rho_0$ does not change at all as time goes on. Instead, the environment denoted by $B$ and $C$ evolve. 
\end{rem}
The easy proof of Theorem \ref{thm:probability_by_dual_process} will be given in the next section.

\section{Proof and Some Analytic Preparation}
In this section we provide with the proof of Theorem \ref{thm:probability_by_dual_process} and we give some analytic preparation which will be useful when we consider asymptotic behavior of some functions. We start with 
\begin{pf}[of Theorem \ref{thm:probability_by_dual_process}]
Recall the inner product in \eqref{eq:H-S_inner_product}. By using Fourier transform and the formula \eqref{eq:evolution_in_Fourier_space}, we see that 
\begin{eqnarray*}
p_x^{(n)}&=&\mathrm{Tr}(\rho_x^{(n)})\\
&=&\left\langle I_2,\rho_x^{(n)}\right\rangle\\
&=&\frac1{2\pi}\int_Ke^{ikx}\left\langle I_2,\widehat{\rho^{(n)}}(k)\right\rangle dk\\
&=&\frac1{2\pi}\int_Ke^{ikx}\left\langle \left(e^{-ik}L_{B^*}R_B+e^{ik}L_{C^*}R_C\right)^n(I_2), \rho_0\right\rangle dk\\
&=&\frac1{2\pi}\int_Ke^{ikx}\mathrm{Tr}\left(\rho_0 \left(e^{ik}L_{B^*}R_B+e^{-ik}L_{C^*}R_C\right)^n(I_2) \right)dk\\
&=& \frac1{2\pi}\int_Ke^{ikx}\mathrm{Tr}\left(\rho_0 Y_n(k)\right)dk.
\end{eqnarray*}
\end{pf} \\
In the next section we will consider several examples and compute the distribution concretely by using Theorem \ref{thm:probability_by_dual_process}. 

Now we introduce some analytic result which is not only interesting in itself but also useful for studying an asymptotic behavior of functions. But, it may be well known in analysis. 
\begin{pro}\label{pro:asymptotics}
Let $[a,b]\subset \RM$ be a finite interval and let $f:[a,b]\to \RM$ be a continuous function such that $|f|$ has a unique maximum at a point $c\in [a,b]$. Then for any continuous $g:[a,b]\to \RM$,
\[
\lim_{n\to \infty} \frac1{\alpha_n}\int_a^bf(x)^ng(x)dx=g(c),
\]
where $\alpha_n=\int_a^bf(x)^ndx$.
\end{pro}
\begin{pf}
By multiplying a constant, if necessary, we may assume $\max_{x\in [a,b]}|f(x)|=f(c)=1$. Also, we may assume $g(x)\ge 0$. We will assume $c\in (a,b)$, because the proof for the case $c=a$ or $b$ is similar. Given a sufficiently small $\epsilon>0$, we can find a $\delta>0$ such that if $|x-c|<\delta$, then $f(x)>0$ and $|g(x)-g(c)|<\epsilon$. Moreover, if $|x-c|\ge \delta$, then $|f(x)|\le 1-\epsilon$. Also, we can find a $0<\delta_1<\delta$ such that if $|x-c|<\delta_1$, then $f(x)\ge 1-\epsilon/2$. First we show that there is an $n_0\in \NM$ such that
\[
\alpha_n>0\quad \text{ if }n\ge n_0.
\]
In fact, we can divide the integral
\begin{equation}\label{eq:division}
\alpha_n=\int_a^bf(x)^ndx=\int_{\{|x-c|<\delta\}}f(x)^ndx+\int_{\{|x-c|\ge \delta\}}f(x)^ndx.
\end{equation}
Notice that 
\begin{eqnarray*}
\int_{\{|x-c|<\delta\}}f(x)^ndx&\ge& \int_{\{|x-c|<\delta_1\}}f(x)^ndx\ge (1-\epsilon/2)^n2\delta_1,\\
\int_{\{|x-c|\ge\delta\}}|f(x)|^ndx&\le &(1-\epsilon)^n(b-a).
\end{eqnarray*}
Thus we see that 
\begin{equation}\label{eq:vanishing}
\lim_{n\to \infty}\frac{\int_{\{|x-c|\ge\delta\}}|f(x)|^ndx}{\int_{\{|x-c|<\delta\}}f(x)^ndx}=0.
\end{equation}
By \eqref{eq:division} and \eqref{eq:vanishing} we have 
\begin{equation}\label{eq:main_part}
\lim_{n\to \infty}\frac1{\alpha_n}\int_{\{|x-c|<\delta\}}f(x)^ndx=1\text{ and }\lim_{n\to \infty}\frac1{\alpha_n}\int_{\{|x-c|\ge\delta\}}|f(x)|^ndx=0,
\end{equation}
and the claim follows. Now notice that
\[
\frac1{\alpha_n}\int_a^bf(x)^ng(x)dx=\frac1{\alpha_n}\int_{\{|x-c|<\delta\}}f(x)^ng(x)dx+\frac1{\alpha_n}\int_{\{|x-c|\ge \delta\}}f(x)^ng(x)dx.
\]
Since $|g(x)|\le M$ for some $M>0$, for $n\ge n_0$
\[
\left|\frac1{\alpha_n}\int_{\{|x-c|\ge \delta\}}f(x)^ng(x)dx\right|\le M\frac1{\alpha_n}\int_{\{|x-c|\ge\delta\}}|f(x)|^ndx.
\]
Hence by \eqref{eq:main_part} we see that 
\begin{equation}\label{eq:minor_part}
\lim_{n\to \infty}\frac1{\alpha_n}\int_{\{|x-c|\ge \delta\}}f(x)^ng(x)dx=0.
\end{equation}
On the other hand, for $n\ge n_0$ we have
\[
\frac{g(c)-\epsilon}{\alpha_n}\int_{\{|x-c|<\delta\}}f(x)^ndx\le \frac1{\alpha_n}\int_{\{|x-c|<\delta\}}f(x)^ng(x)dx\le \frac{g(c)+\epsilon}{\alpha_n}\int_{\{|x-c|<\delta\}}f(x)^ndx.
\]
Therefore, by \eqref{eq:main_part} we see that for sufficiently large $n$'s
\begin{equation}\label{eq:major_part}
\left|\frac1{\alpha_n}\int_{\{|x-c|<\delta\}}f(x)^ng(x)dx-g(c)\right|\le 2\epsilon.
\end{equation}
By \eqref{eq:minor_part} and \eqref{eq:major_part}, the proof is completed.
\end{pf}

\section{Examples}
In this section, in order to see the usefulness of our theorem we consider several examples. We also compare with the result of \cite{AG-PS2012}.  
\subsection{Example 1}
For $p$ and $q$ such that $p+q=1$, $p,q\in [0,1]$, let 
\begin{eqnarray*}
B=
\left[
\begin{array}{cc}
1 & 0 \\
0 & \sqrt{p}
\end{array}
\right],
\qquad
C=
\left[
\begin{array}{cc}
0 & 0 \\
0 & \sqrt{q} 
\end{array}
\right].
\end{eqnarray*}
Since $B$ and $C$ are all diagonal matrices, $L_B$ and $L_C$, $R_B$ and $R_C$ commute. So, it is very easy to compute 
\begin{eqnarray*}
Y_n(k)&=&\left(e^{ik}L_{B^*}R_B+e^{-ik}L_{C^*}R_C\right)^n(I_2)\\
&=&\left(e^{ik}{B^*}B+e^{-ik}{C^*}C\right)^n\\
&=&\sum_{l=0}^n {n\choose l}e^{-ik(n-2l)}\left[\begin{matrix}1&0\\0&p\end{matrix}\right]^l\left[\begin{matrix}0&0\\
0&q\end{matrix}\right]^{n-l}\\
&=&e^{ikn}\left[\begin{matrix}1&0\\0&0\end{matrix}\right]+\sum_{l=0}^ne^{-ik(n-2l)}p^lq^{n-l}\left[\begin{matrix}0&0\\0&1\end{matrix}\right].
\end{eqnarray*}
Therefore, since $\rho_0=\left[\begin{matrix}a&0\\0&b\end{matrix}\right]$, by Theorem \ref{thm:probability_by_dual_process}, we have
\begin{eqnarray*}
p_x^{(n)}&=&\frac1{2\pi}\int_Ke^{ikx}\mathrm{Tr}\left(\rho_0 Y_n(k)\right)dk\\
&=&a\delta_{x,-n}+b\sum_{l=0}^n{n\choose l}p^lq^{n-l}\delta_{x,n-2l}.
\end{eqnarray*}
Thus a standard argument implies
\begin{pro}
Consider OQRW with initial state $\rho_0$. If $p \in (0,1)$, then as $n \to \infty$,
\par
\noindent
(i)
\begin{align*}
\frac{X_n}{n} \quad \Longrightarrow \quad a \> \delta_{-1}+b\>\delta_{q-p},
\end{align*}
\par
\noindent
(ii)
\begin{align*}
\frac{X_n -(q-p)n}{\sqrt{n}} \quad \Longrightarrow \quad b \> N \left(0, 4pq \right).
\end{align*}
\end{pro}
We notice that in this example the equation \eqref{eq:cpm} has infinitely many invariant states.

\subsection{Example 2}
We take 
\begin{eqnarray}\label{eq:generating_matrices}
B=
\left[
\begin{array}{cc}
b_{11} & 0 \\
b_{21} & 0
\end{array}
\right],
\qquad
C=
\left[
\begin{array}{cc}
0 & c_{12} \\
0 & c_{22} 
\end{array}
\right],
\end{eqnarray}
such that $U=B+C \in U(2)$, where $U(2)$ is the set of $2 \times 2$ unitary matrices. We suppose that   
\begin{align*}
p = |b_{11}|^2 = 1 - |b_{21}|^2 = |c_{22}|^2  = 1 - |c_{12}|^2.
\end{align*} 
Let $P_1:=\left[\begin{matrix}1&0\\0&0\end{matrix}\right]$ and $P_2:=\left[\begin{matrix}0&0\\0&1\end{matrix}\right]$ be the projections. From the definition of $B$ and $C$ it is easily shown that
\begin{equation}\label{eq:B_relations}
B^*B=P_1,\quad B^*P_1B=pP_1, \quad B^*P_2B=(1-p)P_1
\end{equation}
and 
\begin{equation}\label{eq:C_relations}
C^*C=P_2,\quad C^*P_1C=(1-p)P_2, \quad C^*P_2C=pP_2.
\end{equation}
Therefore, $Y_n(k)=\left(e^{ik}L_{B^*}R_B+e^{-ik}L_{C^*}R_C\right)^n(I_2)$ becomes a linear combination of $P_1$ and $P_2$.
\begin{lem}\label{lem:dual_process_ex1}
Let 
\begin{equation}\label{eq:linear_combination}
Y_n(k):=a^{(n)}_1(k)P_1+a^{(n)}_2P_2.
\end{equation}
Then the coefficients are determined by
\[
\left[\begin{matrix}a^{(n)}_1(k)\\a^{(n)}_2(k)\end{matrix}\right]=\left[\begin{matrix}e^{ik}p&e^{ik}(1-p)\\
e^{-ik}(1-p)&e^{-ik}p\end{matrix}\right]^n\left[\begin{matrix}1\\1\end{matrix}\right].
\]
\end{lem}
\begin{pf}
We have $Y_{n+1}(k)=\left(e^{ik}L_{B^*}R_B+e^{-ik}L_{C^*}R_C\right)Y_n(k)$. Inserting \eqref{eq:linear_combination}, and by using the relations \eqref{eq:B_relations} and \eqref{eq:C_relations} we get the recurrence relation
\begin{equation}\label{eq:recurrence_ex1}
\left[\begin{matrix}a^{(n+1)}_1(k)\\a^{(n+2)}_2(k)\end{matrix}\right]=\left[\begin{matrix}e^{ik}p&e^{ik}(1-p)\\
e^{-ik}(1-p)&e^{-ik}p\end{matrix}\right]\left[\begin{matrix}a^{(n)}_1(k)\\a^{(n)}_2(k)\end{matrix}\right]. 
\end{equation}
Since $\left[\begin{matrix}a^{(0)}_1(k)\\a^{(0)}_2(k)\end{matrix}\right]=\left[\begin{matrix}1\\1
\end{matrix}\right]$,  the result follows.
\end{pf}
\begin{pro}
The OQRW generated by $B$ and $C$ in \eqref{eq:generating_matrices} is a correlated random walk and as $n \to \infty$, 
\begin{align*}
\frac{X_n}{\sqrt{n}} \quad \Longrightarrow \quad N \left(0, \frac{p}{1-p} \right).
\end{align*} 
\end{pro}
\begin{pf}
By Theorem \ref{thm:probability_by_dual_process} and Lemma \ref{lem:dual_process_ex1}, we have 
\begin{equation}\label{eq:correlated_walk}
p^{(n)}_x=ap^{(n)}_1(x)+bp^{(n)}_2(x),
\end{equation}
where
\[
p^{(n)}_j(x):=\frac1{2\pi}\int_Ke^{ikx}a^{(n)}_j(k)dk,\quad j=1,2.
\]
Since $\left[\begin{matrix}e^{ik}p&e^{ik}(1-p)\\
e^{-ik}(1-p)&e^{-ik}p\end{matrix}\right]=\left[\begin{matrix}e^{ik}&0\\
0&e^{-ik}\end{matrix}\right]\left[\begin{matrix} p& 1-p \\
 1-p & p\end{matrix}\right]$, from \eqref{eq:recurrence_ex1} we get 
\[
\left[\begin{matrix}e^{-ik}&0\\
0&e^{ik}\end{matrix}\right]\left[\begin{matrix}a^{(n+1)}_1(k)\\a^{(n+2)}_2(k)\end{matrix}\right]=\left[\begin{matrix} p& 1-p \\
 1-p & p\end{matrix}\right]\left[\begin{matrix}a^{(n)}_1(k)\\a^{(n)}_2(k)\end{matrix}\right]. 
\]
We multiply $e^{ikx}$ to the both sides of the above equation and integrate w.r.t. $k$. Then we obtain
\[
 \left[\begin{matrix}p^{(n+1)}_1(x-1)\\p^{(n+2)}_2(x+1)\end{matrix}\right]=\left[\begin{matrix} p& 1-p \\
 1-p & p\end{matrix}\right]\left[\begin{matrix}p^{(n)}_1(x)\\p^{(n)}_2(x)\end{matrix}\right], 
\]
or rewriting it we get 
\begin{eqnarray*}
p^{(n+1)}_1(x)&=&pp^{(n)}_1(x+1)+(1-p)p^{(n)}_2(x+1)\\
p^{(n+1)}_2(x)&=&(1-p)p^{(n)}_1(x-1)+pp^{(n)}_2(x-1).
\end{eqnarray*}
From the above relations and \eqref{eq:correlated_walk}, we get the result (see \cite{Konno2009}, for example). \end{pf}

We see that this example falls into the class that the result of \cite{AG-PS2012} may be used. We compute that \eqref{eq:cpm} has a unique invariant state $\rho_\infty=\frac12\left[\begin{matrix}1&(b_{11}{\overline b}_{21}+c_{11}{\overline c}_{22})\\ ({\overline b}_{11}b_{21}+c_{22}{\overline c}_{12})&1\end{matrix}\right]$.

\subsection{Example 3}
Let us define
\begin{eqnarray*}
B=
\left[
\begin{array}{cc}
1 & 0 \\
0 & \sqrt{p - \frac{\gamma^2}{2}}
\end{array}
\right],
\qquad
C=
\left[
\begin{array}{cc}
0 & \gamma \\
0 & \sqrt{q - \frac{\gamma^2}{2}} 
\end{array}
\right],
\end{eqnarray*}
with $p+q=1, \> p,q \in [0,1]$ and $0 < \gamma \le \min \{\sqrt{2p}, \sqrt{2q}\}$. It is promptly shown that
\begin{equation}\label{eq:B_relations_ex3}
B^*B=P_1+\tilde pP_2,\quad B^*P_1B=P_1, \quad B^*P_2B=\tilde pP_2
\end{equation}
and 
\begin{equation}\label{eq:C_relations_ex3}
C^*C=\tilde qP_2,\quad C^*P_1C=\gamma^2P_2, \quad C^*P_2C=\tilde qP_2,
\end{equation}
where $\tilde p=p-\frac{\gamma^2}{2}$ and $\tilde q=q-\frac{\gamma^2}2$.
Therefore, $Y_n(k)=\left(e^{ik}L_{B^*}R_B+e^{-ik}L_{C^*}R_C\right)^n(I_2)$ is again a linear combination of $P_1$ and $P_2$. By denoting $Y_n(k)=a^{(n)}_1(k)P_1+a^{(n)}_2(k)P_2$, we repeat the method done in Lemma \ref{lem:dual_process_ex1} using the relations \eqref{eq:B_relations_ex3} and \eqref{eq:C_relations_ex3}. Then we easily get 
\begin{eqnarray}\label{eq:coefficients_ex3}
\left[\begin{matrix}a^{(n)}_1(k)\\a^{(n)}_2(k)\end{matrix}\right]&=&\left[\begin{matrix}e^{ik}&0\\
e^{-ik}\gamma^2&e^{ik}\tilde p+e^{-ik}\tilde q\end{matrix}\right]^n\left[\begin{matrix}1\\1\end{matrix}\right]\nonumber\\
&=&\left[\begin{matrix}e^{ikn}\\
e^{-ik}\gamma^2\sum_{l=0}^{n-1}e^{ik(n-1-l)}\left(e^{ik}\tilde p+e^{-ik}\tilde q\right)^l+\left(e^{ik}\tilde p+e^{-ik}\tilde q\right)^n\end{matrix}\right].
\end{eqnarray}
\begin{lem}\label{lem:distribution_ex3}
We have 
\begin{align*}
p^{(n)}_x 
&=a \delta_{x, -n} + b \gamma^2 \sum_{j=0}^{n-1} \sum_{l=0}^j {j \choose l} \tilde{p}^l \tilde{q}^{j-l} \delta_{x, 2(j-l+1)-n} 
\\
&+ b \sum_{j=0}^n {n \choose j} \tilde{p}^j \tilde{q}^{n-j} \delta_{x, n-2j}.
\end{align*}
\end{lem}
\begin{pf}
Recall that $Y_n(k)=a^{(n)}_1(k)P_1+a^{(n)}_2(k)P_2$ with $a^{(n)}_j(k)$, $j=1,2$, being defined by \eqref{eq:coefficients_ex3}. By Theorem \ref{thm:probability_by_dual_process} we see that
\begin{equation}\label{eq:density_ex3}
p^{(n)}_x=a\frac1{2\pi}\int_Ke^{ikx}a^{(n)}_1(k)dk+b\frac1{2\pi}\int_Ke^{ikx}a^{(n)}_2(k)dk.
\end{equation}
Inserting the formulas for  $a^{(n)}_j(k)$, $j=1,2$, in \eqref{eq:coefficients_ex3} into \eqref{eq:density_ex3}, we easily get the result.
\end{pf}
\begin{pro}
Consider OQRW with initial state $ \rho_0 = \left[
\begin{array}{cc}
a & 0 \\
0 & b
\end{array}
\right]
\otimes
| 0 \rangle \> \langle 0 |$.  Then 
\begin{align*}
\frac{X_n}{n} \quad \Longrightarrow \quad  \> \delta_{-1}
\end{align*}
as $n\to \infty$.
\end{pro}
\begin{pf} Let us compute the characteristic function $\phi_{X_n/n}(t):=E\left(e^{itX_n/n}\right)$. By using Lemma \ref{lem:distribution_ex3} we get
\begin{align*}
E \left(e^{itX_n/n}\right)
&=\sum_{x\in \ZM}e^{itx/n}P(X_n=x)
\\
&=
ae^{-it}+b\gamma^2e^{-it(n-2)/n} \> \frac{1-\left(\tilde{p}+\tilde{q}e^{2it/n}\right)^n}{1-\left(\tilde{p}+\tilde{q}e^{2it/n}\right)} + be^{it}\left(\tilde{p}e^{-2it/n}+\tilde{q}\right)^n 
\\
& \to e^{-it},
\end{align*}
as $n \to \infty$. In the last line we have used the fact that $\left( \tilde{p}+\tilde{q} e^{2it/n} \right)^n \to 0$ because $\left| \tilde{p}+\tilde{q} e^{2it/n} \right|\le \tilde p+\tilde q<1$. Also similarly $\left( \tilde{p}e^{-2it/n}+\tilde{q} \right)^n \to 0$. Notice that $e^{-it}$ is the characteristic function for the distribution $\delta_{-1}$.
\end{pf}
\begin{rem}\label{rem:ex3}
By the same method as above we can in general show that for all $\alpha>0$ as $n\to \infty$
\[
\frac{X_n+n}{n^\alpha}\Longrightarrow \delta_0.
\]
If we rely on Theorem \ref{thm:clt_AG-PS}, we can show that $m=-1$ and $\sigma^2=0$ and the result says that as $n\to \infty$
\[
\frac{X_n+n}{\sqrt{n}}\Longrightarrow N(0,\sigma^2)
\]
with $\sigma^2=0$.
\end{rem}

\subsection{Example 4}
Here we will consider an example of OQRW whose distribution is a mixture of normal distributions. Let $0<\epsilon$ be a small number such that $2\epsilon a(\epsilon)<1/2$, where $a(\epsilon):=\sqrt{1/2-\epsilon^2}$,  and let $\theta\in \RM$. Define
\begin{eqnarray}\label{eq:matrices_ex4}
B=
\left[
\begin{array}{cc}
a(\epsilon) & \epsilon e^{i\theta} \\
\epsilon e^{i\theta} & a(\epsilon)
\end{array}
\right],
\qquad
C=
\left[
\begin{array}{cc}
a(\epsilon) & -\epsilon e^{i\theta} \\
-\epsilon e^{i\theta} & a(\epsilon) 
\end{array}
\right].
\end{eqnarray}
It is straightforward to see that all the matrices $B, B^*, C$, and $C^*$ commute with each other and 
\begin{eqnarray}\label{eq:matrices_ex4}
B^*B=
\left[
\begin{array}{cc}
1/2 & 2\epsilon a(\epsilon)\cos\theta \\
 2\epsilon a(\epsilon)\cos\theta &1/2
\end{array}
\right],
\qquad
C^*C=
\left[
\begin{array}{cc}
1/2 & -2\epsilon a(\epsilon)\cos\theta \\
 -2\epsilon a(\epsilon)\cos\theta &1/2
\end{array}
\right], 
\end{eqnarray}
and thus
\[
B^*B+C^*C=I_2.
\]
The two matrices $B^*B$ and $C^*C$ are spontaneously diagonalized as 
\begin{eqnarray}
B^*B=U^*\left[
\begin{array}{cc}
\lambda_+(\epsilon,\theta) & 0 \\
0 &\lambda_-(\epsilon,\theta)
\end{array}
\right]U \text{ and } \quad C^*C=U^*\left[
\begin{array}{cc}
\lambda_-(\epsilon,\theta) & 0 \\
0 &\lambda_+(\epsilon,\theta)
\end{array}
\right]U,
\end{eqnarray}
where the eigenvalues are 
\begin{equation}\label{eq:eigenvalues_ex4}
\lambda_{\pm}(\epsilon,\theta)=1/2\pm 2\epsilon a(\epsilon)\cos \theta
\end{equation}
and $U$ is the unitary matrix given by 
\begin{equation}\label{eq:unitary_ex4}
U=\frac1{\sqrt{2}}\left[\begin{array}{cc}
1&1\\1&-1\end{array}\right].
\end{equation}
Because of the commuting properties of the matrices, we can easily compute that
\begin{eqnarray}\label{eq:Y_n(k)_ex4}
Y_n(k)&=&\left(e^{ik}L_{B^*}R_B+e^{-ik}L_{C^*}R_C\right)^n(I_2)\nonumber\\
&=&\left(e^{ik}B^*B+e^{-ik}C^*C\right)^n\nonumber\\
&=&\sum_{l=0}^n {n \choose l} e^{-ik(n-2l)}(B^*B)^l(C^*C)^{n-l}\nonumber\\
&=&\sum_{l=0}^n {n \choose l} e^{-ik(n-2l)}U^*\left[\begin{array}{cc}
\lambda_+^l\lambda_-^{n-l}&0\\
0&\lambda_-^l\lambda_+^{n-l}\end{array}\right]U.
\end{eqnarray}
Now by using Theorem \ref{thm:probability_by_dual_process}, we get the following result.
\begin{pro}\label{prop:ex4_distribution}
Let $B$ and $C$ be the matrices in \eqref{eq:matrices_ex4}, $\lambda_{\pm}\equiv \lambda_{\pm}(\epsilon,\theta)$ the eigenvalues of $B^*B$ and $C^*C$ in \eqref{eq:eigenvalues_ex4}, and let $U$ be the unitary matrix in \eqref{eq:unitary_ex4}. Let $\rho_0$ be the initial state on $\mathcal M_2$. Then the probability distribution of the OQRW defined by $B$ and $C$ is given by 
\[
p^{(n)}_x=a_1p^{(n)}_{1,x}+a_2p^{(n)}_{2,x},
\]
where $a_1=(U\rho_0U^*)_{11}$ and $a_2=(U\rho_0U^*)_{22}$, and $p^{(n)}_{j,x}$, $j=1,2$, are the distributions of random variables $X_{j,n}$, $j=1,2$, respectively, whose asymptotic behavior are as follows: as $n\to \infty$, 
\begin{eqnarray*}
&&\frac{X_{1,n}-(\lambda_- -\lambda_+)n}{\sqrt{n}}\Rightarrow 
N(0,4\lambda_+ \lambda_- ),\\
&&\frac{X_{2,n}-(\lambda_+ -\lambda_- )n}{\sqrt{n}}\Rightarrow 
N(0,4\lambda_+ \lambda_- ).
\end{eqnarray*}
\end{pro}
\begin{pf}
From \eqref{eq:Y_n(k)_ex4}, we see that 
\[
\mathrm{Tr}(\rho_0Y_n(k))=\sum_{l=0}^n{n\choose l}e^{-ik(n-2l)}(a_1\lambda_+^l\lambda_-^{n-l}+a_2\lambda_-^l\lambda_+^{n-l}),
\]
where $a_1=(U\rho_0U^*)_{11}$ and $a_2=(U\rho_0U^*)_{22}$.
Therefore by Theorem \ref{thm:probability_by_dual_process}, we have
\[
p^{(n)}_x=\sum_{l=0}^n{n\choose l}(a_1\lambda_+^l\lambda_-^{n-l}+a_2\lambda_-^l\lambda_+^{n-l})\delta_{x,n-2l}.
\]
Now the result follows from the standard arguments.
\end{pf}

\subsection{Example 5}
Define 
\begin{eqnarray}\label{eq:matrices_ex5}
B=
\frac{1}{\sqrt{3}}
\left[
\begin{array}{cc}
1 & 1 \\
0 & 1
\end{array}
\right],
\qquad
C=
\frac{1}{{\sqrt3}}
\left[
\begin{array}{cc}
1 & 0 \\
-1 & 1 
\end{array}
\right].
\end{eqnarray}
We let 
\begin{eqnarray}\label{eq:Y_n(k)_ex5}
Y_n(k)&=&\left(e^{ik}L_{B^*}R_B+e^{-ik}L_{C^*}R_C\right)^n(I_2)\nonumber\\
&=&\left[\begin{matrix}a^{(n)}_{11}(k)&a^{(n)}_{12}(k)\\
a^{(n)}_{21}(k)&a^{(n)}_{22}(k)\end{matrix}\right].
\end{eqnarray}
By directly computing we get the recursion relation:
\[
\left[\begin{matrix}a^{(n+1)}_{11}(k)\\
a^{(n+1)}_{12}(k)\\
a^{(n+1)}_{21}(k)\\
a^{(n+1)}_{22}(k)\end{matrix}\right]
= Y(k) \left[\begin{matrix}a^{(n)}_{11}(k)\\
a^{(n)}_{12}(k)\\
a^{(n)}_{21}(k)\\
a^{(n)}_{22}(k)\end{matrix}\right]
\]
where
\begin{equation}
Y(k):=\frac{1}{3}
\left[
\begin{array}{cccc}
2 \cos k & -e^{-ik} & -e^{-ik} & e^{-ik} \\
e^{ik} & 2 \cos k & 0 & -e^{-ik} \\
e^{ik} & 0 & 2 \cos k & -e^{-ik} \\
e^{ik} & e^{ik} & e^{ik} & 2 \cos k
\end{array}
\right].
\end{equation}
Thus we have the solution
\begin{equation}\label{eq:form_of_Y_n(k)_ex5}
\left[\begin{matrix}a^{(n)}_{11}(k)\\
a^{(n)}_{12}(k)\\
a^{(n)}_{21}(k)\\
a^{(n)}_{22}(k)\end{matrix}\right]
=Y(k)^n\left[\begin{matrix}1\\0\\0\\1\end{matrix}\right].
\end{equation}
By Theorem \ref{thm:probability_by_dual_process} we have 
\begin{eqnarray}\label{eq:density_ex5}
p^{(n)}_x&=&\frac1{2\pi}\int_Ke^{ikx}\mathrm{Tr}(\rho_0Y_n(k))dk\nonumber\\
 &=&\frac1{2\pi}\int_Ke^{ikx}\left(aa^{(n)}_{11}(k)+ba^{(n)}_{22}(k)\right)dk.
\end{eqnarray}

We can introduce a combinatoric way to compute the distribution. Notice that 
\begin{equation}\label{eq:powers}
{B^*}^nB^n=\frac1{3^n}\left[\begin{matrix}1&n\\n&n^2+1\end{matrix}\right]\quad \mathrm{ and }\quad
{C^*}^nC^n=\frac1{3^n}\left[\begin{matrix}n^2+1&-n\\-n&1\end{matrix}\right].
\end{equation}
Thus 
\begin{equation}\label{eq:power_sum}
{B^*}^nB^n+{C^*}^nC^n=\frac{n^2+2}{3^n}I_2.
\end{equation}
We notice also that 
\[
\mathrm{Tr}{B^*}^nB^n=\mathrm{Tr}{C^*}^nC^n=\frac{n^2+2}{3^n}.
\]
Since $Y_n(k)=\left(e^{ik}L_{B^*}R_B+e^{-ik}L_{C^*}R_C\right)^n(I_2)$, by expanding the power and using Theorem \ref{thm:probability_by_dual_process}, we see that $p^{(n)}_x$ is the sum of all the contributions from the terms of the type
\begin{eqnarray}\label{eq:contributor}
&&(L_{C^*}R_C)^{l_s}(L_{B^*}R_B)^{r_s}(L_{C^*}R_C)^{l_{s-1}}(L_{B^*}R_B)^{r_{s-1}}\cdots(L_{C^*}R_C)^{l_1}(L_{B^*}R_B)^{r_1}(I_2)\nonumber \\
&&={C^*}^{l_s}{B^*}^{r_s}{C^*}^{l_{s-1}}{B^*}^{r_{s-1}}\cdots {C^*}^{l_1}{B^*}^{r_1}B^{r_1}C^{l_1}\cdots B^{r_{s-1}}C^{l_{s-1}}B^{r_s}C^{l_s},
\end{eqnarray}
where $\sum_{j=1}^s(l_j-r_j)=x$ and $r_1\ge 0$, $r_j\ge 1$, $j=2,\cdots, s$, $l_j\ge 1$, $j=1,\cdots, s-1$, and  $l_s\ge 0$. We would like to expand \eqref{eq:contributor}. By \eqref{eq:power_sum}, the term ${B^*}^{r_1}B^{r_1}$ in the middle is equal to $\frac{r_1^2+2}{3^{r_1}}I_2-{C^*}^{r_1}C^{r_1}$. Inserting this into \eqref{eq:contributor}, we get a sum of two sequences, whose middle terms are ${C^*}^{l_1}C^{l_1}$ multiplied by a factor $\frac{r_1^2+2}{3^{r_1}}$ and ${C^*}^{(l_1+r_1)}C^{(l_1+r_1)}$ multiplied by a factor $-1$, respectively. We continue this process successively. For it, it is very convenient to understand \eqref{eq:contributor} as a random walk path (assume all $r_j$'s and $l_j$'s are greater than $0$ for simplicity): the walker goes upward $r_1$ units, then goes $l_1$ units downward. Then it goes $r_2$ units upward and $l_2$ units downward, and so on. So, the path is a continuously connected lines consisting of $2s$ segments (which have different lengths of $r_1$, $l_1$, etc.). We will further simplify the notation by denoting it just as a sequence $(l_s,r_s,\cdots, l_1,r_1)$. We will make short the sequence step by step by applying the process of "cutting" or "unfolding". For example, at first step, if we make cutting we will get the sequence $(l_s,r_s,\cdots,l_1)$ with a weight $\frac{r_1^2+2}{3^{r_1}}$. Instead, if we make unfolding at the first step, we get a new sequence $(l_s,r_s,\cdots, l_1+r_1)$ with weight $-1$. Any operation shortens the sequence by length $1$. We continue this process until we get a length one sequence, or for the random walk path, until it remains a single segment of length $l_s+l'$, say. The resultant matrix is nothing but ${C^*}^{l_s+l'}C^{l_s+l'}$ and we need to compute the trace $\mathrm{Tr}(\rho_0{C^*}^{l_s+l'}C^{l_s+l'})$, which is simply $\frac1{3^{l_s+l'}}\left(a\left((l_s+l')^2+1\right)+b\right)$. We summarize this process as a theorem. Below $\sum_{l_1,\cdots,l_s, r_1,\cdots, r_s}^{(x)}$ means the sum over sequences such that $\sum_{j=1}^s(l_j-r_j)=x$ and $r_1\ge 0$, $r_j\ge 1$, $j=2,\cdots, s$, $l_j\ge 1$, $j=1,\cdots, s-1$, and  $l_s\ge 0$. $\mathcal C\mathcal U(l_s,r_s,\cdots, l_1,r_1)$ means the set of all sequences of shortening process of cutting and unfolding upto a single term and for $\pi\in \mathcal C\mathcal U(l_s,r_s,\cdots, l_1,r_1) $, $l(\pi)$ is the length of the remaining single segment for the process $\pi$, and $\omega(\pi)$ is the product of the weights of $\pi$
\begin{thm}\label{thm:distribution_ex5}
The probability distribution for the OQRW determined by $B$ and $C$ in \eqref{eq:matrices_ex5} is given as follows:
\[
p^{(n)}_x=\sum_{ l_1,\cdots,l_s, r_1,\cdots, r_s}^{(x)}\sum_{\pi\in \mathcal C\mathcal U(l_s,r_s,\cdots,l_1,r_1)}\omega(\pi)\mathrm{Tr}\left(\rho_0{{\overline C}^*}^{l(\pi)}{\overline C}^{l(\pi)}\right),
\]
where ${\overline C}=C$ if $l_s\neq 0$ and ${\overline C }=B$ if $l_s=0$.
\end{thm}
As an example, let us compute the distribution of $X_4$ for the case $a=b=1/2$. Since $\mathrm{Tr}({B^*}^lB^l)=\mathrm{Tr}({C^*}^lC^l)$ for all $l\ge 0$, it is easy to see that the distribution under the assumption is symmetric. So, we only need to compute $P(X_4=4)$ and $P(X_4=2)$. The random walk path leading to $X_4=4$ is a single segment consisting of $4$ upward units. Or, in the symbol of finite sequence, it is just $(r_1)=(4)$. Thus, we get 
\[
P(X_4=4)=\mathrm{Tr}(\rho_0{B^*}^4B^4)=\frac12\frac{4^2+2}{3^4}=\frac19.
\]
For $X_4=2$, we have $4$-paths: $(-+++), (+-++), (++-+), (+++-)$, or in symbols of sequences
$(\overset  {l_1} {1},\overset{r_1}3)$, $(\overset{r_2}1, \overset{l_1}1,\overset{r_1}2)$, $(\overset{r_2}2,\overset{l_1}1,\overset{r_1}1)$, and $(\overset{r_2}3,\overset{l_1}1)$, respectively. For each symbol we apply cutting-unfolding process.
\begin{eqnarray*}
(1,3):&&\frac{3^2+2}{3^3}\mathrm{Tr}(\rho_0C^*C)-\mathrm{Tr}(\rho_0{C^*}^4C^4)=\frac5{54}\\
(1,1,2):&&\frac{2^2+2}{3^2}(1,1)-(1,3)\\
&&=\frac{2^2+2}{3^2}\left(\frac{1^2+2}{3^1}\mathrm{Tr}(\rho_0B^*B)-\mathrm{Tr}(\rho_0{B^*}^2B^2)\right)-\left(\frac{3^2+2}{3^3}\mathrm{Tr}(\rho_0B^*B)-\mathrm{Tr}(\rho_0{B^*}^4B^4)\right)\\
&&=\frac1{54}\\
(2,1,1):&&(2,1)-(2,2)\\
&&=\left(\mathrm{Tr}(\rho_0{B^*}^2B^2)-\mathrm{Tr}(\rho_0{B^*}^3B^3)\right)-\left(\frac{2^2+2}{3^2}\mathrm{Tr}(\rho_0{B^*}^2B^2)-\mathrm{Tr}(\rho_0{B^*}^4B^4)\right)=\frac1{54}\\
(3,1):&&\mathrm{Tr}(\rho_0{B^*}^3B^3)-\mathrm{Tr}(\rho_0{B^*}^4B^4)=\frac{5}{54}.
\end{eqnarray*}
Thus summing all the contributions we get $P(X_4=2)=\frac29$. Using the symmetry, we see that $\mu_4$,  the distribution of $X_4$, is equal to 
\[
\mu_4=\frac19\delta_{-4}+\frac29\delta_{-2}+\frac39\delta_{0}+\frac29\delta_{2}+\frac19\delta_{4}.
\]

From now on we discuss the asymptotic behavior of the distribution. Recall the matrix $Y_n(k)$ in \eqref{eq:Y_n(k)_ex5} and its representation in \eqref{eq:form_of_Y_n(k)_ex5}. The eigen-equation of $Y(k)$ is 
\begin{align*}
\left(\lambda -\frac{2\cos k}{3}\right)\left(\lambda^3 -2 \cos k \lambda^2 + \frac{4 \cos^2 k+1}{3} \lambda - \frac{2 \cos k (4 \cos^2 k +5)}{27}\right)=0,
\end{align*}
so the eigenvalues of $Y(k)$ are 
\begin{align}\label{eq:eigenvalues1_ex5}
\lambda_0 
&= \frac{2 \cos k}{3}, \quad
\lambda_1 
= \frac{2 \cos k}{3} + \frac{1}{3} \left( \xi - \frac{1}{\xi} \right),
\\ \label{eq:eigenvalues2_ex5}
\lambda_2 
&= \frac{2 \cos k}{3} + \frac{1}{6} \left\{ \left( -1 + i \sqrt{3} \right) \xi + \frac{1 + i \sqrt{3}}{\xi} \right\}, \quad \lambda_3 = \overline{\lambda_2},
\end{align}
where
\begin{align}\label{eq:function_xi}
\xi = \xi (k) = \left( 2 \cos k + \sqrt{4 \cos^2 k+1} \right)^{1/3}.
\end{align}
Let
\begin{align*}
A_1 = \lambda_1 - \lambda_0, \quad  A_2= \lambda_2 - \lambda_0, \quad  A_3= \lambda_3 - \lambda_0.
\end{align*}
Then a direct computation gives
\begin{align*}
Y(k) = S^{-1} \> \mathrm{diag} [\lambda_0, \lambda_1, \lambda_2, \lambda_3] \> S,
\end{align*}
where $S:=R^T$ with
\begin{align*}
R
=
\frac{1}{6}
\left[
\begin{array}{cccc}
0 & 2 - 6 e^{ik} A_1 & 2 - 6 e^{ik} A_2 & 2 - 6 e^{ik} A_3 \\
6 & 1 - 9  A_1^2 & 1 - 9   A_2^2 & 1 - 9   A_3^2 \\
-6 & 1 - 9   A_1^2 & 1 - 9   A_2^2 & 1 - 9   A_3^2 \\
0 & -2 + 6 e^{-ik} A_1 & -2 + 6 e^{-ik} A_2 & -2 + 6 e^{-ik} A_3
\end{array}
\right],
\end{align*}
and $\mathrm{diag} [a_0, a_1, \ldots , a_n]$ denotes the diagonal matrix whose $(i,i)$-component is $a_{i}$. Here
\begin{align*}
R^{-1}
=
\left[
\begin{array}{cccc}
0 & 1/2 & -1/2 & 0 \\
w_{21} & w_{22} & w_{23} & w_{24} \\
w_{31} & w_{32} & w_{33} & w_{34} \\
w_{41} & w_{42} & w_{43} & w_{44}
\end{array}
\right].
\end{align*}
Recall the density formula in \eqref{eq:density_ex5}. We need to compute $a^{(n)}_{jj}(k)$, $k=1,2$, which we can obtain from the diagonalization of $Y(k)$. We note that $\det R = 4 \sqrt{3(4 \cos^2 k +1)} \> \sin k/9.$  After a little computation we have 
\begin{eqnarray}\label{eq:integrand_ex5}
p^{(n)}_x&=&\frac1{2\pi}\int_{-\pi}^{\pi} e^{ikx}\left(aa^{(n)}_{11}(k)+ba^{(n)}_{22}(k)\right)dk\nonumber\\ 
&=& \frac1{2\pi}\int_{-\pi}^\pi e^{ikx}\left(- 2 i \sin k \> \sum_{j=1}^3 \> A_j \> (w_{j+1,1} a + w_{j+1,4} b ) \> \lambda_j^n\right)dk.
\end{eqnarray}
In order to get an information of the asymptotic behavior or $p^{(n)}_x$ as $n\to \infty$, we need to investigate the eigenvalues more carefully, and then we will rely on Proposition \ref{pro:asymptotics}.  For that purpose we will rewrite the eigenvalues. Recall the eigenvalues in \eqref{eq:eigenvalues1_ex5} and  \eqref{eq:eigenvalues2_ex5}, and the function $\xi(k)$ in \eqref{eq:function_xi}. Since $\cos k$ appears in the eigenvalues, we let $u:=\cos k$. Then $u$ varies in the interval $[-1,1]$ and we have
\[
\xi=\xi(u)= (2u+\sqrt{4u^2+1})^{1/3}.
\]
Further, we define 
\[
s=s(u):=\xi(u)-\frac1{\xi(u)}, \quad -1\le u\le 1.
\]
It is not hard to show that 
\begin{eqnarray*}
\xi(-1)&=& (-2+\sqrt{5})^{1/3}=\frac{\sqrt{5}-1}2,\\
\xi(1)&=&(2+\sqrt{5})^{1/3}=\frac{\sqrt{5}+1}2.
\end{eqnarray*}
Moreover, on the interval $[-1,1]$, the function $s(u)$ is increasing with
\[
s(-1)=-1 \text{ and } s(1)=1.
\]
We can also check that 
\[
\xi^3-\frac1{\xi^3}=4u.
\]
Therefore, the eigenvalues can be rewritten as 
\begin{eqnarray*}
\lambda_0&=&\frac16(\xi^3-\frac1{\xi^3})=\frac16 s(s^2+3),\\
\lambda_1&=&\frac16s(s^2+5)\\
\lambda_2&=&\frac16s(s^2+2)+i\frac{\sqrt{3}}6\sqrt{s^2+4},\\
\lambda_3&=&\overline {\lambda_2}.
\end{eqnarray*}
Since $-1\le s\le 1$, we see that 
\begin{equation}\label{eq:eigenvalues_upper_bound}
|\lambda_0|\le 2/3,\quad |\lambda_2|\le \sqrt{2/3}.
\end{equation}
Only the eigenvalue $\lambda_1$ moves fully on the interval $[-1,1]$: $\lambda_1(u=-1)=-1$, $\lambda_1(u=1)=1$. Regarding $K:=(-\pi,\pi]$ as a unit circle in the plane, as usual, it is not hard to see that  the eigenvalue $\lambda_1\equiv \lambda_1(k)$ is anti-symmetric in the sense that $\lambda_1(k+\pi)=-\lambda_1(k)$. Moreover, it behaves very much similar to the cosine function. In particular, $\lambda_1(k)\ge 0 $ on $[-\pi/2,\pi/2]$ and it is negative on $K\setminus [-\pi/2,\pi/2]$.  Let us define a scaling constant $\alpha_n$ by 
\begin{equation}\label{eq:scaling_ex5}
\alpha_n:=\int_{-\pi/2}^{\pi/2}\lambda_1(k)^ndk.
\end{equation}
\begin{lem}\label{lem:asymptotics_ex5}
For $j=0,1,2,3$, let $g_j(k)$ be continuous functions on $K$. Then 
\[
\lim_{n\to \infty}\frac1{\alpha_n}\int_K\sum_{j\neq 1} g_j(k)\lambda_j(k)^ndk=0,
\]
and 
\begin{eqnarray*}
&&\lim_{n\to \infty}\frac1{\alpha_{2n}}\int_K g_1(k)\lambda_1(k)^{2n}dk= g_1(0)+g_1(\pi),\\
&&\lim_{n\to \infty}\frac1{\alpha_{2n-1}}\int_K g_1(k)\lambda_1(k)^{2n-1}dk= g_1(0)-g_1(\pi).
\end{eqnarray*}
\end{lem}
\begin{pf}
First, since $\lambda_1(k)$ is continuous and  $\lambda_1(0)=1$, as in the proof of Proposition \ref{pro:asymptotics}, it is very easy to see that for any $\sqrt{2/3}<q<1$, 
\begin{equation}\label{eq:scale_lower_bound}
\lim_{n\to \infty}\frac{q^n}{\alpha_n}=0.
\end{equation}
On the other hand, by \eqref{eq:eigenvalues_upper_bound} 
\[
\left|\int_K\sum_{j\neq 1} g_j(k)\lambda_j(k)^ndk\right|=O((2/3)^{n/2}).
\] 
From this and by \eqref{eq:scale_lower_bound}, the first assertion follows. For the second assertion, we divide the integral:
\[
\frac1{\alpha_{n}}\int_K g_1(k)\lambda_1(k)^{n}dk=\frac1{\alpha_{n}}\int_{[-\pi/2,\pi/2]} g_1(k)\lambda_1(k)^{n}dk+\frac1{\alpha_{n}}\int_{K\setminus [-\pi/2,\pi/2]} g_1(k)\lambda_1(k)^{n}dk.
\]
Noticing the anti-symmetry of $\lambda_1(k)$, i.e., $\lambda_1(k+\pi)=-\lambda_1(k)$, the result follows from Proposition \ref{pro:asymptotics}.
\end{pf}\\
Looking at the formula \eqref{eq:integrand_ex5}, by Lemma \ref{lem:asymptotics_ex5}, we see that asymptotically the term containing $\lambda_1^n$ dominates. By direct computation we have
\begin{eqnarray*}
\omega_{21}&=&\frac{(A_2-A_3)}{\mathrm{det}R}\left(\frac13e^{-ik}
-(A_2+A_3)+3e^{-ik}A_2A_3\right)\\
\omega_{24}&=&\frac{(A_2-A_3)}{\mathrm{det}R}\left(\frac13e^{ik}
-(A_2+A_3)+3e^{ik}A_2A_3\right)
\end{eqnarray*}
Now we can get the proper asymptotics for the density $p_x^{(n)}$.
\begin{thm}\label{thm:aymptotics_ex5}
As $n\to \infty$, the asymptotic behavior of $p_x^{(n)}$ is as follows.
\begin{eqnarray*}
\lim_{n\to \infty}\frac{p_x^{(2n)}}{\alpha_{2n}}&=&\begin{cases}1/\pi, &\textrm{if  }x\textrm{ is even},\\
0, &\textrm{if } x \textrm{ is odd}.\end{cases}\\
\lim_{n\to \infty}\frac{p_x^{(2n-1)}}{\alpha_{2n-1}}&=&
 \begin{cases}0, &\textrm{if  }x\textrm{ is even},\\
1/\pi, &\textrm{if } x \textrm{ is odd}.\end{cases}.
\end{eqnarray*}
\end{thm}
\begin{pf} The proof will follow from Lemma \ref{lem:asymptotics_ex5} with 
\[
g_1(k)=C(k)(aB_1(k)+bB_2(k)),
\]
where 
\begin{eqnarray*}
C(k)&=&\frac{-9ie^{ikx}(A_2-A_3)}{4\pi \sqrt{3(4\cos^2k+1)}},\\
B_1(k)&=&A_1\left(\frac13e^{-ik}
-(A_2+A_3)+3e^{-ik}A_2A_3\right),\\
B_2(k)&=&A_1\left(\frac13e^{ik}
-(A_2+A_3)+3e^{ik}A_2A_3\right).
\end{eqnarray*}
We need to know the values $g_1(0)$ and $g_1(\pi)$. Let us define a symbol $\eta$ by 
\[
\eta(0):=1, \quad \eta(\pi):=-1.
\]
By directly computing, we get
\[
C(0)=\frac3{4\pi}\quad \mathrm{ and }\quad C(\pi)=\frac3{4\pi}\cos \pi x.
\]
Also it is easy to check that when restricted to $\{0,\pi\}$, $A_1=\frac13\eta$ and the quantities in the brackets $(\cdots)$ of $B_1$ and $B_2$ are equal to $2\eta$. Thus we see that when restricted to $\{0,\pi\}$ 
\[
B_1=B_2=2/3.
\]
Combining these we use Lemma \ref{lem:asymptotics_ex5} to get the result.  
\end{pf}
 
Concerning the central limit theorem of this example we have the following result. 
\begin{thm}
For the example of this subsection, we have as $n\to \infty$
\[
\frac{X_n}{\sqrt{n}}\Longrightarrow N(0,8/9).
\]
\end{thm}
\begin{pf}  Let us consider the characteristic function
\[
\EM[e^{itX_n/\sqrt{n}}]=\sum_{x\in \ZM}e^{itx/\sqrt{n}}p_x^{(n)}.
\]
Here, $p_x^{(n)}$ is given in \eqref{eq:integrand_ex5}, but since the eigenvalue $\lambda_1$ dominates we have
\[
p_x^{(n)}\sim \int_{-\pi}^\pi e^{ikx}f^{(n)}(k)dk,
\]
where
\[
f^{(n)}(k)=g(k)\lambda_1(k)^n
\]
with 
\[
g(k)=\frac1{12\pi}\frac1{\sqrt{4\cos^2k+1}}\left(\xi^2-\frac1{\xi^2}\right)\left\{\cos k\left(\xi+\frac1{\xi}\right)^2+\xi-\frac1{\xi}\right\}.
\]
By putting $y=x/\sqrt{n}$ and taking a change of variable  $m=\sqrt{n}k$ we have
\[
\EM[e^{itX_n/\sqrt{n}}]\sim \sum_{y\in \ZM/\sqrt{n}}e^{ity}  \frac1{\sqrt{n}}\int_{-\sqrt{n}\pi}^{\sqrt{n}\pi}e^{imy}g({m}/{\sqrt{n}})\l_1({m}/{\sqrt{n}})^n \,{dm}.  
\] 
Notice that the function $g(k)$ is bounded and continuous and $\lambda_1(k)$ behaves very much similar to cosine function on the interval $[-\pi, \pi]$. Now we expand the interval to $[-\sqrt{n}\pi,\sqrt{n}\pi]$ and take a power $n$ to $\lambda_1$. As $n$ grows, the function $\lambda_1(m/\sqrt{n})^n$, when integrated with a multiplication by a mild function $g(m/\sqrt{n})$, will pick up the values of $g(m/{\sqrt{n}})$ at $m=0$ and $m=\pm \sqrt{n}\pi$ with dominating factors of itself. Notice that $g(0)=g(\pm\pi)=\frac1{2\pi}$. We first consider the behavior at $m=0$. For that we notice that $\lambda_1(k)$ has a Taylor expansion at $k=0$ as 
\[
\lambda_1(k)=1-\frac49k^2+\frac1{54}k^4+o(k^4).
\]
Therefore, 
\[
\lambda_1(m/\sqrt{n})^n\sim e^{-\frac49m^2}.
\]
Thus, the contribution to $\EM[e^{itX_n/\sqrt{n}}]$ is evaluated as
\begin{eqnarray}{\label{eq:gaussian}}
&&\sum_{y\in \ZM/\sqrt{n}}e^{ity}  \frac1{\sqrt{n}}\frac1{2\pi}\int_{-\infty}^\infty e^{imy}e^{-\frac49m^2}dm\nonumber\\
&=&\sum_{y\in \ZM/\sqrt{n}}e^{ity}  \frac1{\sqrt{n}}\frac1{2\pi}\frac32\sqrt{\pi}e^{-\frac{9}{16}y^2}\nonumber\\
&\to&\frac1{\sqrt{2\pi\sigma^2}}\int_{-\infty}^\infty e^{ity}e^{-\frac1{2\sigma^2}y^2}dy\quad \text{as }n\to \infty,
\end{eqnarray}
 where $\sigma^2=8/9$. 
Next we consider the effect coming from the factor $\lambda_1(k)^n$ at $k=\pm \pi$. For this, it is convenient to shift the integration interval as
\[
p_x^{(n)}\sim \int_{0}^{2\pi} e^{ikx}f^{(n)}(k)dk.
\]
Now $\lambda_1(\pi)=-1$ and by a similar argument as above the contribution to $\EM[e^{itX_n/\sqrt{n}}]$ is 
\begin{eqnarray*}
&&\sum_{y\in \ZM/\sqrt{n}}e^{ity}  \frac1{\sqrt{n}}\frac1{2\pi}(-1)^n\int_{-\infty}^\infty e^{imy}e^{-\frac49(m-\sqrt{n}\pi)^2}dm\\
&=& (-1)^n\sum_{y\in \ZM/\sqrt{n}}e^{ity}  \frac1{\sqrt{n}}\frac1{2\pi} e^{i\sqrt{n}\pi y}\frac32\sqrt{\pi}e^{-\frac{9}{16}y^2}.
\end{eqnarray*}
Now as $n\to \infty$, by an argument of Riemann-Lebesgue Lemma, the last term converges to 0. Combining this with \eqref{eq:gaussian}, we conclude that in the example 4, 
\[
\frac{X_n}{\sqrt{n}}\Rightarrow N(0, 8/9).
\]

\end{pf}

This example was also dealt with in \cite{AG-PS2012} too. There they computed the invariant state of \eqref{eq:cpm} obtaining $\rho_\infty=\frac12I$. They also computed the mean $m=0$ and variance $\sigma^2=\frac89$, the same result as we obtained here. 
\par
\
\par\noindent
{\bf Acknowledgment.} This work was partially supported by the Grant-in-Aid for Scientific Research (C) of Japan Society for the Promotion of Science (Grant No. 21540116).
\par
\
\par
\begin{small}
\bibliographystyle{jplain}

\end{small}

\end{document}